\documentclass[11pt,a4paper]{article}
\usepackage[margin=2.5cm]{geometry}
\usepackage{amsmath, amsfonts, amssymb, graphicx, caption, subcaption}
\usepackage{hyperref}
\usepackage{siunitx}
\usepackage{natbib}
\usepackage{charter}
\usepackage{setspace}
\usepackage{float}

\title{\textsc{\Large From data to decisions: Bayesian modelling and global sensitivity analysis for flotation control}}

\author{
Paulina Quintanilla \\
Department of Chemical Engineering\\
University College London\\
London, United Kingdom \\
\texttt{p.quintanilla@ucl.ac.uk} \\
\And
Agustín Fuenzalida \\
Departamento de Ingeniería Química y Ambiental\\
Universidad Técnica Federico Santa María\\
Santiago, Chile \\
\texttt{agustin.fuenzalida@usm.cl} \\
\And
Daniel Navia \\
Departamento de Ingeniería Química y Ambiental\\
Universidad Técnica Federico Santa María\\
Santiago, Chile \\
\texttt{daniel.navia@usm.cl} \\
\And
Pablo Brito-Parada \\
Department of Earth Science and Engineering\\
Imperial College London\\
London, United Kingdom \\
\texttt{p.brito-parada@imperial.ac.uk} \\
}

\makeatletter
\renewcommand{\maketitle}{%
  \begin{center}
    \vspace*{1ex}
    \rule{\textwidth}{0.8pt}\par
    \vspace{1ex}
    {\LARGE\bfseries \MakeUppercase{\@title}\par}
    \vspace{1ex}
    \rule{\textwidth}{0.8pt}\par
  \end{center}

  \vspace{1.5em}

  \noindent
  \begin{minipage}[t]{0.48\textwidth}\centering
    \textbf{Paulina Quintanilla}\par
    {\small
    Department of Chemical Engineering\par
    University College London\par
    London, United Kingdom\par
    }\texttt{p.quintanilla@ucl.ac.uk}
  \end{minipage}\hfill
  \begin{minipage}[t]{0.48\textwidth}\centering
    \textbf{Agustín Fuenzalida}\par
    {\small
    Departamento de Ingeniería Química y Ambiental\par
    Universidad Técnica Federico Santa María\par
    Santiago, Chile\par
    }\texttt{agustin.fuenzalida@usm.cl}
  \end{minipage}

  \vspace{1.8em}

  \noindent
  \begin{minipage}[t]{0.48\textwidth}\centering
    \textbf{Daniel Navia}\par
    {\small
    Departamento de Ingeniería Química y Ambiental\par
    Universidad Técnica Federico Santa María\par
    Santiago, Chile\par
    }\texttt{daniel.navia@usm.cl}
  \end{minipage}\hfill
  \begin{minipage}[t]{0.48\textwidth}\centering
    \textbf{Pablo Brito-Parada}\par
    {\small
    Department of Earth Science and Engineering\par
    Imperial College London\par
    London, United Kingdom\par
    }\texttt{p.brito-parada@imperial.ac.uk}
  \end{minipage}

  \vspace{2em}
}
\makeatother

\begin{document}

\maketitle

\begin{abstract}
This work presents a data-driven framework for interpretable modelling and decision support in flotation systems, integrating Gaussian Process (GP) regression with Global Sensitivity Analysis (GSA) via Sobol indices and local interpretability using SHapley Additive exPlanations (SHAP). Based on laboratory-scale experimental data, a static GP surrogate model is developed to capture how superficial air velocity, overflowing froth velocity, froth height over the lip, pulp height, bubble size, and tailings flowrate influence the measured air recovery. The trained GP enables the computation of Sobol indices to quantify the contribution of each variable and their interactions to the overall variance in air recovery. The combination of Bayesian inference and Sobol-based sensitivity metrics provides a systematic approach to identify the dominant and interacting variables governing air recovery. This study links Bayesian learning, sensitivity quantification, and explainability to provide a foundation for data-driven control and optimisation of flotation processes.
\end{abstract}

\section{Introduction}


The increasing availability of process data offers new opportunities for advanced modelling and optimisation. Recent developments in Bayesian machine learning provide a principled framework to integrate data and uncertainty in process modelling. Gaussian Process (GP) regression, in particular, offers a nonparametric, probabilistic approach capable of capturing complex input–output relationships while maintaining interpretability through kernel structure and uncertainty propagation \citep{Rasmussen2006}. At the same time, techniques such as Global Sensitivity Analysis (GSA) \citep{sobol2001global, saltelli2008} and SHapley Additive exPlanations (SHAP) \citep{lundberg2017unified} enable the quantification and visualisation of variable influence and interactions, both globally and locally. The integration of these methods can produce interpretable data-driven models for mineral processes, enabling informed decision-making for control and optimisation.

This work focuses on air recovery ($\alpha$) as the target output variable, due to its established role as a proxy for froth stability and overall flotation efficiency. The aim is to construct a Bayesian surrogate model that captures $\alpha$ dynamics under varying operating conditions, quantify the global and local importance of the input variables, and interpret the resulting insights within the context of flotation control. 

\section{Gaussian Process regression for air recovery modelling}

GP models define a distribution over functions, offering both a mean prediction and a measure of uncertainty at every input location. Let $\boldsymbol{x} \in \mathbb{R}^d$ denote an input vector comprising $d$ normalised process variables, and $y \in \mathbb{R}$ represent the observed air recovery. The prior over the latent function $f(\boldsymbol{x})$ is defined as a Gaussian Process:
\begin{equation}
f(\boldsymbol{x}) \sim \mathcal{GP}(m(\boldsymbol{x}), k(\boldsymbol{x}, \boldsymbol{x}')),
\end{equation}
where $m(\boldsymbol{x})$ is the mean function (typically zero) and $k(\boldsymbol{x}, \boldsymbol{x}')$ is the covariance kernel. In this work, we employ a squared exponential kernel with Automatic Relevance Determination (ARD):
\begin{equation}
k(\boldsymbol{x}, \boldsymbol{x}') = \sigma_f^2 \exp \left( -\frac{1}{2} \sum_{i=1}^{d} \frac{(x_i - x_i')^2}{\ell_i^2} \right),
\end{equation}
where $\ell_i$ represents the characteristic length scale of the $i$-th input, $\sigma_f^2$ is the signal variance, and $\sigma_n^2$ is the noise variance.

Given $n$ training samples $\boldsymbol{X} = [\boldsymbol{x}_1, \ldots, \boldsymbol{x}_n]^T$ and observations $\boldsymbol{y} = [y_1, \ldots, y_n]^T$, the joint distribution of training and test outputs is:
\begin{equation}
\begin{bmatrix}
\boldsymbol{y} \\
f_* 
\end{bmatrix}
\sim
\mathcal{N}\left( 
\boldsymbol{0}, 
\begin{bmatrix}
K(\boldsymbol{X}, \boldsymbol{X}) + \sigma_n^2 I & K(\boldsymbol{X}, \boldsymbol{x}_*) \\
K(\boldsymbol{x}_*, \boldsymbol{X}) & K(\boldsymbol{x}_*, \boldsymbol{x}_*)
\end{bmatrix}
\right),
\end{equation}
where $K(\boldsymbol{X}, \boldsymbol{X})$ is the covariance matrix evaluated at the training points. The predictive mean and variance at a new test point $\boldsymbol{x}_*$ are given by:
\begin{align}
\mu(\boldsymbol{x}_*) &= K(\boldsymbol{x}_*, \boldsymbol{X}) [K(\boldsymbol{X}, \boldsymbol{X}) + \sigma_n^2 I]^{-1} \boldsymbol{y}, \\
\sigma^2(\boldsymbol{x}_*) &= K(\boldsymbol{x}_*, \boldsymbol{x}_*) - K(\boldsymbol{x}_*, \boldsymbol{X}) [K(\boldsymbol{X}, \boldsymbol{X}) + \sigma_n^2 I]^{-1} K(\boldsymbol{X}, \boldsymbol{x}_*).
\end{align}

The model hyperparameters $\boldsymbol{\theta} = \{\ell_i, \sigma_f, \sigma_n\}$ are obtained by maximising the log marginal likelihood:
\begin{equation}
\log p(\boldsymbol{y} | \boldsymbol{X}, \boldsymbol{\theta}) = 
- \frac{1}{2} \boldsymbol{y}^T [K(\boldsymbol{X}, \boldsymbol{X}) + \sigma_n^2 I]^{-1} \boldsymbol{y}
- \frac{1}{2} \log |K(\boldsymbol{X}, \boldsymbol{X}) + \sigma_n^2 I| - \frac{n}{2} \log 2\pi.
\end{equation}

This formulation provides both a best-estimate prediction $\mu(\boldsymbol{x}_*)$ and a credible interval given by $\mu(\boldsymbol{x}_*) \pm 1.96\,\sigma(\boldsymbol{x}_*)$, corresponding to the 95\% confidence level. The probabilistic nature of GP makes it well suited for risk-aware control and optimisation, as predictive uncertainty can be directly propagated through decision-making layers such as model predictive control.

\section{Global sensitivity analysis}

To interpret the trained surrogate model and quantify how input variables influence air recovery, we apply a variance-based Global Sensitivity Analysis (GSA) following the Sobol framework \citep{sobol2001global}. The GSA decomposes the total output variance into contributions from each input and their interactions.

Let $y = f(\boldsymbol{x})$ be the model output, with $\boldsymbol{x} = [x_1, x_2, \ldots, x_d]$ defined over the unit hypercube. The total variance of $y$ can be expressed as:
\begin{equation}
V(y) = \mathrm{Var}[f(\boldsymbol{x})] = \sum_{i=1}^d V_i + \sum_{i<j} V_{ij} + \ldots + V_{12\ldots d},
\end{equation}
where $V_i$ is the variance contribution of input $x_i$ alone, $V_{ij}$ represents the interaction between $x_i$ and $x_j$, and higher-order terms capture more complex interactions.

The Sobol indices are defined as:
\begin{align}
S_i &= \frac{V_i}{V(y)}, \\
S_{ij} &= \frac{V_{ij}}{V(y)}, \\
S_{T_i} &= 1 - \frac{V_{\sim i}}{V(y)},
\end{align}
where $S_i$ represents the first-order sensitivity index, $S_{ij}$ the second-order index, and $S_{T_i}$ the total-order index that includes all interactions involving $x_i$. These indices satisfy $\sum_i S_i \le 1$ and provide a quantitative ranking of input importance.

In practice, the variances are estimated through Monte Carlo integration using random or quasi-random (Sobol sequence) sampling. For each input, two independent sampling matrices $A$ and $B$ are generated, and hybrid matrices $A_B^{(i)}$ are constructed by replacing the $i$-th column of $A$ with that of $B$. The first-order and total-order indices can then be estimated as:
\begin{align}
S_i &= \frac{\mathrm{Cov}[f(B), f(A_B^{(i)})]}{V(y)}, \\
S_{T_i} &= 1 - \frac{\mathrm{Cov}[f(A), f(A_B^{(i)})]}{V(y)}.
\end{align}

The GSA was implemented using the \texttt{SALib} Python package, evaluating the GP surrogate across 10,000 Monte Carlo samples spanning the normalised range of each input variable.

\section{Data and model validation}

The datasets used in this study were obtained from controlled experimental campaigns on a laboratory flotation tank described in \citet{quintanilla2021dynamic2}. Two time series were analysed, each containing twelve operating conditions characterised by distinct combinations of gas superficial velocity ($j_g$), overflowing froth velocity ($v_f$), froth height over the lip ($h_{over})$, pulp height ($h_p$), tailings flowrate, and bubble size ($d_{32}$). Each condition produced a time series of air recovery, smoothed and normalised prior to modelling.

The trained GP surrogate demonstrated strong predictive performance across all operating conditions. The predicted and measured $\alpha$ curves exhibited close agreement, with credibl tintervals reflecting model uncertainty. In transient regions where data were sparse, the predicted uncertainty widened appropriately, reflecting Bayesian caution rather than overconfidence. The temporal error between predicted and measured $\alpha$ values remained low, with a mean absolute error below 0.5\% for most conditions. This validation confirms the GP model as an accurate and probabilistic digital representation of the underlying process dynamics, suitable for subsequent sensitivity and explainability analysis.

\section{Results and Discussion}

\subsection{Model validation and prediction accuracy}

\begin{figure}[htbp]
    \centering
    \includegraphics[width=0.95\textwidth]{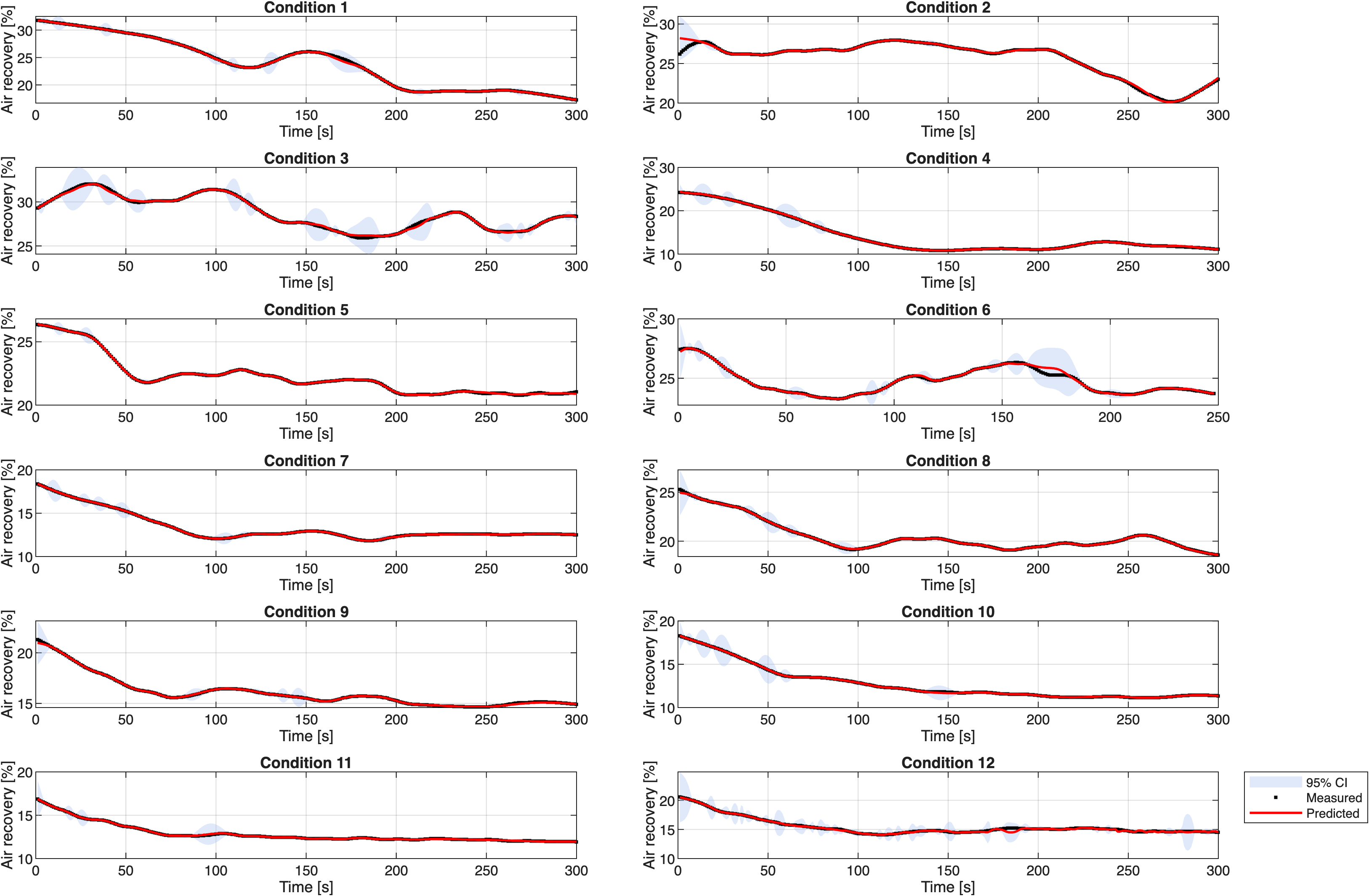}
    \caption{Measured (black) and predicted (red) air recovery trajectories with 95\% confidence intervals (blue shading) for twelve operating conditions.}
    \label{fig:AR_all_R3}
\end{figure}

The trained GP surrogate accurately reproduced the air recovery dynamics across all twelve operating conditions. Figure~\ref{fig:AR_all_R3} compares the measured and predicted $\alpha$ trajectories, showing close agreement throughout both the transient and steady-state regions. The shaded areas correspond to the 95\% confidence intervals, which appropriately widen in regions of higher uncertainty or sparse data, reflecting the Bayesian nature of the model. 

\begin{figure}[htbp]
    \centering
    \includegraphics[width=0.95\textwidth]{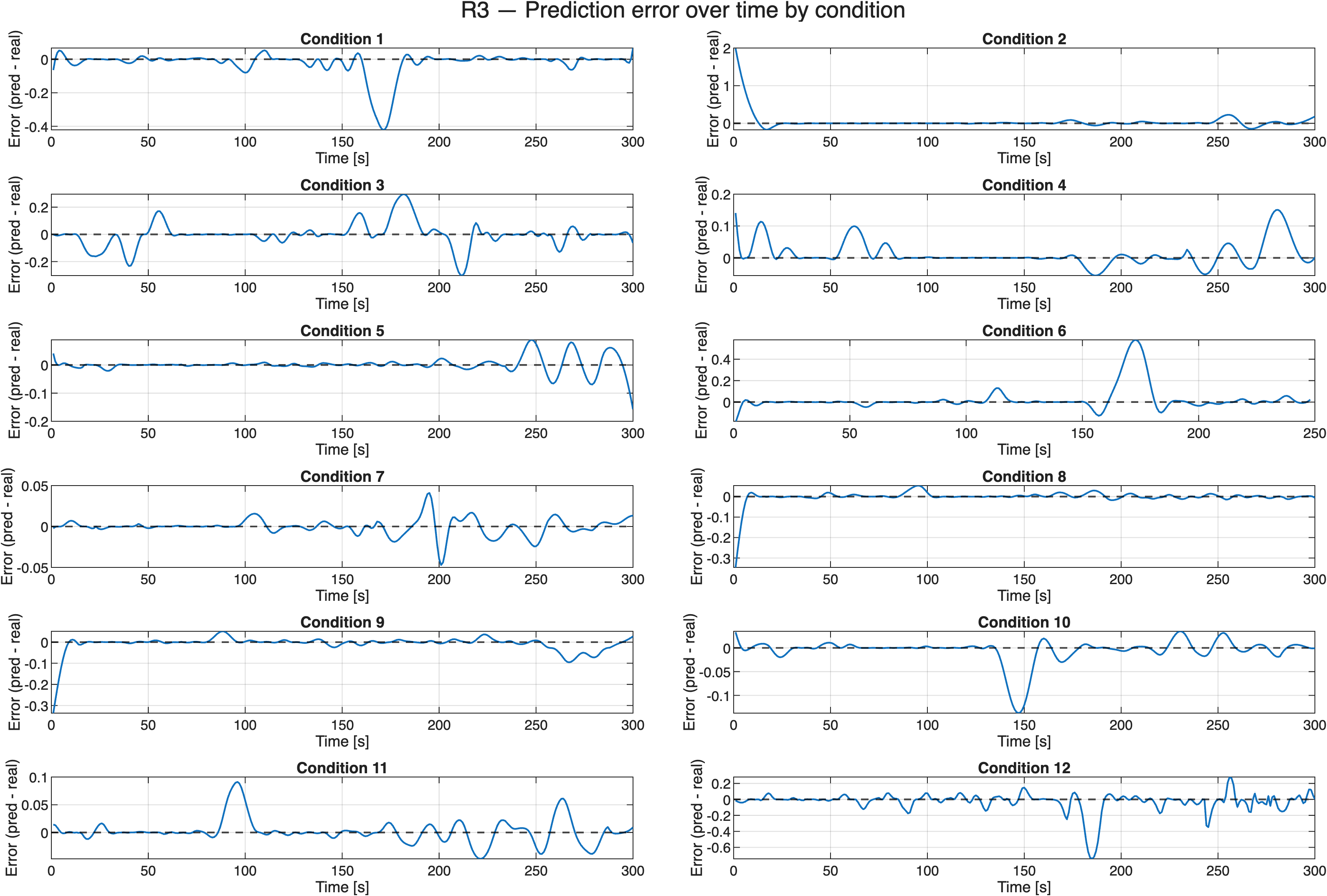}
    \caption{Prediction error ($\text{$\alpha$}_{\text{pred}} - \text{$\alpha$}_{\text{real}}$) over time for twelve operating conditions.}
    \label{fig:Error_all_R3}
\end{figure}

\begin{figure}[htbp]
    \centering
    \includegraphics[width=0.6\textwidth]{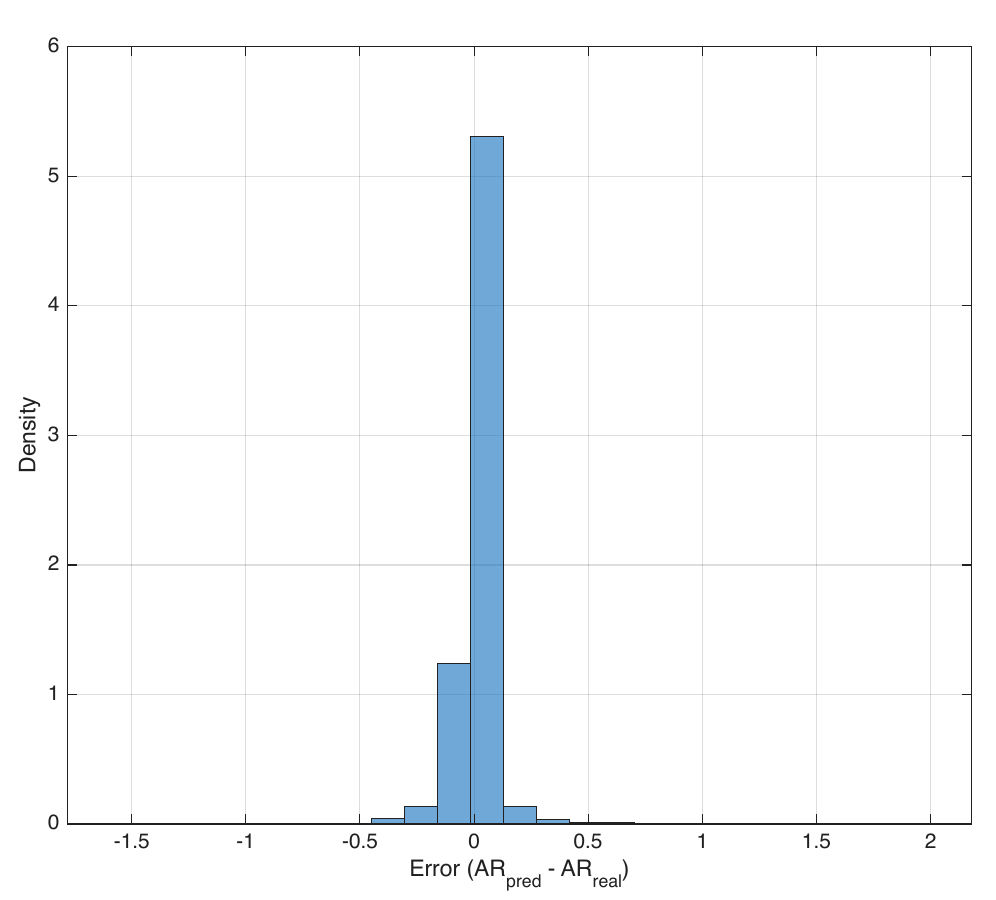}
    \caption{Distribution of prediction errors across all test conditions. The distribution is approximately centred around zero, indicating unbiased model performance.}
    \label{fig:PredictionErrors}
\end{figure}

The corresponding prediction errors, shown in Figure~\ref{fig:Error_all_R3}, remained centred around zero with limited dispersion, confirming the absence of systematic bias. Only isolated deviations were observed under conditions of sharp transients or regime shifts, which are inherently more difficult to capture due to the limited number of samples in those regions. The model exhibited mean absolute errors below 0.5\% for most conditions, and the error distribution (Figure~\ref{fig:PredictionErrors}) showed a narrow, approximately Gaussian profile with zero mean. 

\subsection{Global sensitivity analysis}

Global sensitivity analysis using Sobol indices revealed that $j_g$ and $v_f$ were the dominant variables influencing $\alpha$. $h_{over}$ and $d_{32}$ showed secondary but non-negligible effects. At the same time, pulp height and tailings flowrate contributed minimally to the output variance. 
The total-order Sobol indices highlighted strong nonlinear interactions between $j_g$ and $d_{32}$, confirming their coupled role in controlling bubble generation and gas dispersion. Interactions between $v_f$ and froth height were also present.

\subsection{Local interpretability through SHAP analysis}

While Sobol indices provide global importance rankings, SHapley Additive exPlanations (SHAP) enable local interpretability of model predictions. The results show that $j_g$ had the largest average influence on $\alpha$, followed by $v_f$ and froth height, in agreement with the global sensitivity trends. However, SHAP values additionally expose directionality and nonlinear effects. Higher $j_g$ values were consistently associated with positive SHAP contributions (i.e. increased $\alpha$), whereas higher $v_f$ values showed a negative impact.

\section{From interpretability to control: GP-based NMPC}

Recent advances in data-driven control have demonstrated that GP models can be integrated into nonlinear model predictive control (NMPC) formulations, enabling predictive control of systems with partial observability and uncertainty \citep{wang2025gaussian}. The resulting GP-NMPC framework formulates an optimisation problem of the form:

\begin{equation}
\begin{aligned}
\min_{\boldsymbol{u}_{0:N}} \quad & J(\boldsymbol{z}_{1:N}, \boldsymbol{u}_{0:N-1}, \boldsymbol{\sigma}_{1:N}) \\
\text{subject to:} \quad 
& \boldsymbol{\xi}_k = [\boldsymbol{\mu}_k^T, \boldsymbol{u}_k^T]^T, \\
& \boldsymbol{z}_{k+1} = \mathcal{J}_z(\boldsymbol{\mu}_{k+1}), \\
& \boldsymbol{\mu}_{k+1}, \boldsymbol{\sigma}_{k+1} \leftarrow \boldsymbol{f}_{\Psi}(\boldsymbol{\mu}_{k,\text{obs}}, \boldsymbol{u}_k), \\
& \boldsymbol{\mu}_{0,\text{obs}} = \boldsymbol{x}_\text{obs}, \\
& \boldsymbol{u}_{\text{min}} \leq \boldsymbol{u}_k \leq \boldsymbol{u}_{\text{max}}, \\
& \text{for } k = 0, \dots, N-1.
\end{aligned}
\end{equation}

Here, $\boldsymbol{\xi}_k$ is the concatenated vector of the predicted state mean $\boldsymbol{\mu}_k$ and control input $\boldsymbol{u}_k$ at time step $k$, while $N$ denotes the control horizon. The variable $\boldsymbol{z}_k = \mathcal{J}_z(\boldsymbol{\mu}_k)$ represents an economic or performance-related quantity used in the objective function. The GP model, denoted by \( \mathcal{GP}(m_{\psi}(\boldsymbol{\xi}_k), k_{\psi}(\boldsymbol{\xi}_k, \boldsymbol{\xi}_k')) \), predicts both the mean $\boldsymbol{\mu}_{k+1}$ and the variance $\boldsymbol{\sigma}_{k+1}$ of the process output. The bounds \( \boldsymbol{u}_{\text{min}} \) and \( \boldsymbol{u}_{\text{max}} \) constrain the feasible input space.

This formulation incorporates predictive uncertainty through either Bayesian risk terms or chance constraints, allowing for risk-aware and adaptive control actions. The interpretability framework developed in this work can be directly embedded into the control design. GSA provides quantitative guidance on variable prioritisation. For example, significant nonlinear interactions (e.g. between $j_g$ and $d_{32}$) identified through SHAP analysis can be explicitly accounted for in multivariable constraints or coordination terms within the NMPC structure. 

\section{Conclusions}

This study presents an integrated Bayesian framework for interpretable modelling and decision support in froth flotation. The Gaussian Process surrogate accurately captures the dynamic behaviour of air recovery under diverse operating conditions, while quantifying predictive uncertainty. Global Sensitivity Analysis identifies the dominant process variables and interactions, confirming the physical relevance of gas and liquid fluxes as primary control levers. SHAP analysis complements this by providing local interpretability, linking individual predictions to their causal drivers. Together, these methods provide a transparent bridge between data, process understanding, and control-oriented decision-making.



\small

\bibliographystyle{apalike}
\bibliography{cas-refs}

@book{Rasmussen2006,
  author =	 {Rasmussen, Carl Edward and Williams, Christopher
                  K. I.},
  title =	 {{Gaussian Processes for Machine Learning}},
  year =	 {2016},
  publisher =	 {MIT Press},
  ISBN =	 {0-262-18253-X}
}

@book{saltelli2008,
  author = {Saltelli, A.},
  description = {Global sensitivity analysis: the primer - Andrea Saltelli - Google Bücher},
  isbn = {9780470059975},
  lccn = {2007045551},
  publisher = {John Wiley},
  title = {Global sensitivity analysis: the primer},
  url = {http://books.google.at/books?id=wAssmt2vumgC},
  year = 2008
}

@article{lundberg2017unified,
  title={A unified approach to interpreting model predictions},
  author={Lundberg, Scott M and Lee, Su-In},
  journal={Advances in neural information processing systems},
  volume={30},
  year={2017}
}

@article{sobol2001global,
  title={Global sensitivity indices for nonlinear mathematical models and their Monte Carlo estimates},
  author={Sobol, Ilya M},
  journal={Mathematics and computers in simulation},
  volume={55},
  number={1-3},
  pages={271--280},
  year={2001},
  publisher={Elsevier}
}

@article{quintanilla2021dynamic2,
  title={A dynamic flotation model for predictive control incorporating froth physics. Part II: Model calibration and validation},
  author={Quintanilla, Paulina and Neethling, Stephen J and Mesa, Diego and Navia, Daniel and Brito-Parada, Pablo R},
  journal={Minerals Engineering},
  doi = {10.1016/j.mineng.2021.107190},
  volume={173},
  pages={107190},
  year={2021},
  publisher={Elsevier}
}

@article{wang2025gaussian,
  title={Gaussian Process Nonlinear Model Predictive Control for Online Partially Observable Systems: An Application to Froth Flotation},
  author={Wang, Yicong and del Rio Chanona, Ehecatl Antonio and Quintanilla, Paulina},
  journal={Industrial \& Engineering Chemistry Research},
  year={2025},
  publisher={ACS Publications}
}

\end{document}